\documentclass[twocolumn,showpacs,preprintnumbers,amsmath,amssymb,nofootinbib,APS]{revtex4}

\usepackage{dcolumn}
\usepackage{bm}
\usepackage[latin1]{inputenc}
\usepackage[spanish,english]{babel}
\usepackage{amsfonts}
\usepackage{amssymb}
\usepackage{graphicx}

\newcommand{\be}{\begin{equation}}
\newcommand{\ee}{\end{equation}}
\newcommand{\bea}{\begin{eqnarray}}
\newcommand{\eea}{\end{eqnarray}}

\begin{document}

\title{{\bf The Power Spectrum in de Sitter Inflation, Revisited}}
\author{Iván Agulló and José Navarro-Salas}\email{ivan.agullo@uv.es  ,  jnavarro@ific.uv.es}
\affiliation{ {\footnotesize Departamento de Física Teórica and
IFIC, Centro Mixto Universidad de Valencia-CSIC.
    Facultad de Física, Universidad de Valencia,
        Burjassot-46100, Valencia, Spain. }}

\author{Gonzalo J. Olmo}\email{golmo@perimeterinstitute.ca}
\affiliation{ {\footnotesize Perimeter Institute for Theoretical
Physics, Waterloo, Ontario, N2L 2Y5 Canada}\\ and \\Instituto de Estructura de la Materia, CSIC, Serrano 121, 28006 Madrid, Spain}

\author{Leonard Parker}\email{leonard@uwm.edu}
\affiliation{ {\footnotesize Physics Department, University of
Wisconsin-Milwaukee, P.O.Box 413, Milwaukee, WI 53201 USA}}

\date{May 30, 2008}

\begin{abstract}
We find that the amplitude of quantum fluctuations of the invariant
de Sitter vacuum coincides exactly with that of the vacuum of a
comoving observer for a massless scalar (inflaton) field. We propose
redefining the actual physical power spectrum as the  difference
between the amplitudes of the above vacua. An inertial particle detector
continues to observe the Gibbons-Hawking temperature. However,
although the resulting power spectrum is still scale-free, its  amplitude can
be drastically reduced since now, instead of the Hubble's scale at the
inflationary period, it is determined by the square of the mass of the inflaton
fluctuation field.

\end{abstract}


\maketitle

 The prediction of a  nearly ``scale-free''
spectrum of density perturbations is commonly considered as a
crucial prediction of inflationary cosmology \cite{inflation}.
Departures from homogeneity arise then as quantum fluctuations,
$\phi$, of the scalar inflaton field that drives inflation
\cite{inflation2} (see also \cite{mukhanov-feldman-brandenberger}).
This prediction explains the power spectrum of the galaxy
distribution and has also been successfully confirmed by high
precision measurements \cite{WMAP} of the anisotropies in the cosmic
microwave background.
The amplitude of the spectrum was predicted to be proportional to
the square of the Hubble constant during inflation
 \be\label{amplitudeH2}
\Delta^2_{\phi}(k) \approx \hbar H^2 \ , \ee although the precise
estimate depends on the details of particular models. The resulting
amplitude for GUT-scale inflation turned out to be
several orders of magnitude too large, or required
fine-tuning for model parameters, to account for the observed
$\delta \rho/\rho\sim 10^{-4}-10^{-5}$ and is still a rather elusive
problem.

A simple argument that gives the above amplitude estimate comes from
the Gibbons-Hawking radiation effect.  As measured by a particle
detector on a geodesic, the invariant vacuum state $|0_{dS}\rangle$
in de Sitter space \cite{bunch-davies78} has a non-zero temperature,
the Gibbons-Hawking temperature $T_{GH}= \frac{H\hbar}{2\pi}$
\cite{gibbons-hawking77}, where $H$ is the Hubble constant of the
exponentially expanding de Sitter universe \be \label{desitter} ds^2
= -dt^2 + a^2(t)d\vec{x}^2 \ , \ee with $a(t)= e^{Ht}$.  A comoving
observer detects a thermal bath of radiation at temperature $T_{GH}$
and the associated amplitude of thermal fluctuations accounts for
(\ref{amplitudeH2}).

 However, in deriving the Gibbons-Hawking effect
one implicitly considers two vacuum states. In addition to the
globally defined de Sitter vacuum, we have the local vacuum
$|0_C\rangle$ associated with a freely falling, or comoving,
observer at a given spatial point $\vec{x}$. The comoving observer
perceives the de Sitter vacuum as a thermal bath of particles, with
respect to the $|0_C\rangle$ vacuum. The fundamental argument
underlying this result, as first explained for a general $a(t)$ in
\cite{parker69} (see also \cite{birrel-davies, parker-toms}), is
that the modes defined in the two different vacua are related by a
superposition of positive and negative frequencies and the
corresponding creation and annihilation operators by a Bogoliubov
transformation.  In the present case, the positive-frequency modes
defining the comoving vacuum $u^{C}_i$ cannot be expressed in terms
of the purely positive-frequency modes $u^{dS}_j$ defining the de
Sitter vacuum. It is just the comparison of one set of modes with
respect to the other set that precludes the physical equivalence of
both vacua, and the existence of an horizon for the comoving
observer is then responsible for the exact thermal behavior of
$|0_{dS}\rangle$ in the Fock space of $|0_C\rangle$. The fundamental
role of both vacua can  be nicely displayed in terms of two-point
functions. The standard formula for the expectation value of the
particle number operator in terms of Bogoliubov coefficients
\cite{parker69} can be rewritten, when particularized to de Sitter
space and the inhomogeneous scalar field $\phi$, as follows
\cite{agullo-navarro-salas-olmo-parker}
 \bea \label{a+anormalordering}
\langle 0_{dS}|N_{i}^{C}|0_{dS}\rangle = \sum_j|\beta_{ij}|^2 =
 \frac{1}{\hbar}\int_\Sigma d\Sigma_1 ^\mu d\Sigma_2 ^\nu
(u^{C}_{i}(x_1){\buildrel\leftrightarrow\over{\partial}}_\mu
)\times \\
 (u^{C*}_{i}(x_2){\buildrel\leftrightarrow\over{\partial}}_\nu )
[\langle 0_{dS}| \phi (x_1)\phi (x_2)|0_{dS}\rangle - \langle 0_{C}|
\phi (x_1)\phi (x_2)|0_{C}\rangle] \ \nonumber , \eea where $\Sigma$ is a Cauchy hypersurface. Explicit evaluation of
the above expressions, either via Bogolubov coefficients
\cite{brandenberger} or two-point functions, reproduces the
Planckian spectrum. The physical idea in the latter method
is that it is just the difference between the correlations of the de
Sitter vacuum and those of the comoving vacuum that produces the
relevant observables. Similarly, in black hole emission
\cite{hawk1}, the difference between the two-point function
for the ``in'' vacuum, defined at the remote past before
gravitational collapse, and that for the ``out'' vacuum, defined at
future infinity, is at the heart of Hawking radiation
\cite{agullo-navarro-salas-olmo-parker}. This idea can be reinforced
by deriving the Gibbons-Hawking effect in terms of the Unruh particle
detector \cite{unruh1}. The rate of the response function of an
inertial detector in de Sitter space, with trajectory
$x^{\mu}=x^{\mu}(\tau)$, is given by \be \label{rateFGH}{\dot F(w)}=
\int_{-\infty}^{+\infty}d\Delta \tau e^{-iw\Delta \tau}\langle
0_{dS}|\phi(x(\tau))\phi(x(\tau + \Delta \tau)) |0_{dS}\rangle \ ,
\ee which reproduces, via a detailed balance argument, the expected
thermal result at the temperature $T_{GH}$ \cite{gibbons-hawking77}.
Since the response function of the comoving detector vanishes
in the comoving vacuum, using $i\epsilon$ prescription,
\be \label{rateC}\int_{-\infty}^{+\infty}d\Delta \tau e^{-iw\Delta
\tau}\langle 0_{C}|\phi(x(\tau))\phi(x(\tau + \Delta \tau))
|0_{C}\rangle =0 \ , \ee one can, equivalently, compute the rate
(\ref{rateFGH}) by subtracting the corresponding two-point function
of the comoving observer \cite{agullo-navarro-salas-olmo-parker08}
\begin{eqnarray}\label{rate}
\dot{F}(w)&=& \int_{-\infty}^{+\infty}d\Delta \tau e^{-iw\Delta
\tau} [\langle 0_{dS}|\phi(x(\tau))\phi(x(\tau + \Delta \tau))
|0_{dS}\rangle  \nonumber\\ &-&  \langle
0_C|\phi(x(\tau))\phi(x(\tau + \Delta \tau)) |0_C\rangle]  \ .
\end{eqnarray}
The $i\epsilon$ regularization prescription of the Wightman function in
(\ref{rateFGH})   can be replaced, as a mathematical identity, by the
subtraction of the two-point function for the comoving vacuum. Note that
the integrand in (\ref{rate}) is now a smooth function as a consequence of the
Hadamard condition for the two-point functions \cite{waldbook} and there is no
need for the $i\epsilon$ prescription.
Expression (\ref{rate}) shows again that the detector responds
to the relative correlations between the quantum state and the
vacuum of the comoving observer.

Having in mind  all the above,  we find it natural to propose that, to
properly quantify the amplitude of quantum fluctuations, one should
compare the amplitude of the modes $u^{dS}_i$ of the invariant de Sitter
vacuum with respect to the amplitude of the modes $u^C_j$ of a
comoving observer. This leads us to replace the standard definition of
the power spectrum \cite{mukhanov-feldman-brandenberger,
LiddleLyth2000, Dodelson2003, mukhanovbook} \be
\label{definition1}  \int_0^{\infty} \frac{dk}{k}
\Delta^2_{\phi}(k, t; dS)=\langle 0_{dS}|\phi(t, \vec{x})\phi(t,
\vec{x})|0_{dS} \rangle  \ , \ee
by the following
\begin{eqnarray}
\label{definition2} \int_0^{\infty} \frac{dk}{k}
\Delta^2_{\phi}(k, t)& \equiv &\langle 0_{dS}|\phi(t, \vec{x})\phi(t,
\vec{x})|0_{dS} \rangle \nonumber \\ &-&\langle 0_C| \phi(t, \vec{x})\phi(t,
\vec{x})|0_C\rangle \ .
\end{eqnarray}
In this letter we explore the consequences of this proposal. An
advantage of the new definition (\ref{definition2}) is that its
right-hand side is again a smooth function as a consequence of the
Hadamard condition. If we expand the right-hand side of
(\ref{definition2}) in modes, the integrand is finite and no further
renormalization is needed for $\Delta^2_{\phi}(k,t)$.
With the standard definition (\ref{definition1}) the
right-hand side is formally divergent, implying that renormalization
may play an important role in the evaluation of the physical power
spectrum, as suggested and studied in \cite{parker07}. Nevertheless,
it should be clear that the reason for subtracting the amplitude of
the comoving observer is more fundamental than simply to bypass the
divergence of the two-point functions at the coincident point. The
subtraction would be natural even if there were no divergences.

Let us consider a
minimally coupled scalar field in de Sitter space with $[\Box -
(m/\hbar)^2]\phi(x) =0 $, where $\phi$ can be
thought of as the quantum fluctuation of the inflaton field,
$\phi_0(t)+\phi(x)$, and $m$ is the mass of $\phi(x)$.
The normalized modes $u^{dS}_{\vec{k}}(\vec{x},
t)$ for the invariant de Sitter vacuum are \be
u^{dS}_{\vec{k}}(\vec{x}, t) = \frac{1}{\sqrt{2(2\pi)^3
a(t)^3}}h_k(t) e^{i\vec{k}\vec{x}} \ , \ee \be h_k(t) =
\sqrt{\frac{\pi}{2H}}H^{(1)}_n(kH^{-1}\exp(-Ht)) \ , \ee where
$n=\sqrt{9/4 - m^2/H^2\hbar^2}$ is the index of the Hankel function.
Therefore, the amplitude of quantum fluctuations is given as a sum
in modes \be \langle 0_{dS}|\phi(t, \vec{x})\phi(t, \vec{x})|0_{dS}
\rangle =  \hbar (4\pi^2 a(t)^3)^{-1}\int_0^{\infty} |h_k(t)|^2 k^2 dk
\ , \ee and the standard power spectrum is given by \be
\Delta^2_{\phi}(k, t; dS)= \hbar (4\pi^2 a(t)^3)^{-1}k^3|h_k(t)|^2 \ .
\ee Evaluated in terms of the physical comoving wavevector $\bar{k}=
k / a(t)$ the amplitude behaves as in Minkowski space for very large
$\bar{k}$, but around the exit from the Hubble horizon
$\bar{k}\approx H$, and for $m \ll H \hbar$, one gets the usual
nearly scale-free spectrum \be \Delta^2_{\phi}(\bar{k}; dS) =
\frac{\hbar H^2}{8\pi}|H^{(1)}_n(\bar{k}H^{-1})|^2 \approx
\frac{\hbar H^2}{2\pi}\ . \ee

Let us now study the amplitude of the comoving modes at a given
spatial point $\vec{x}$. To this end it is convenient to introduce
static spherical coordinates \be ds^2 = -(1-
H^2\tilde{r}^2)d\tilde{t}^2 + \frac{d\tilde{r}^2}{1-H^2\tilde{r}^2}
+ \tilde{r}^2 d\Omega^2 \ , \ee where, as usual, $d\vec{x}^{2}=
dr^2+ r^2d\Omega^2$, and  $\tilde{t} = -(2H)^{-1}\ln [e^{-2tH} -
(rH)^2]$, $\tilde{r} = e^{tH}r$. We locate the origin of radial
coordinates $\tilde{r}=0$ at the location $\vec{x}$ of the arbitrary
comoving observer. Note that, at $\tilde{r}=0$, the new time
coordinate $\tilde{t}$ coincides with the comoving time $t$, the
metric takes the Minkowskian form, and the deviations from it are
quadratic in $\tilde{r}$. In evaluating the amplitude of
fluctuations $\langle 0_C| \phi(t, \vec{x})\phi(t,
\vec{x})|0_C\rangle$ at the origin of coordinates only the $s$-wave
sector contributes, due to the regularity condition at
$\tilde{r}=0$. The $s$-modes are found to be \be u^C_{w}=
\frac{e^{-iw\tilde{t}}}{\sqrt{4\pi}} \frac{N_n(w)}{\tilde{r}}
[P^{i\frac{w}{H}}_{n-1/2} (H \tilde{r})- \alpha_n (w)
Q^{i\frac{w}{H}}_{n-1/2}(H \tilde{r})] \ee where $P_\nu^\mu(z)$ and
$Q_\nu^\mu(z)$ are generalized Legendre functions, $N_n(w)$ is a
normalization constant and $\alpha_n(w)$ is a constant ensuring the
regularity at $\tilde{r}=0$ \be \alpha_n(w)= \frac{1}{\pi} \left[ 2
i+ 4 \left(i + e^{- i \pi n} e^{\pi w/H}\right)^{-1} \right] \ee A
major technical point is to compute the exact form of the
normalization constant. Evaluating the scalar product at the future
horizon, with tortoise coordinate $x\equiv
H^{-1}\tanh^{-1}(\tilde{r}H)$, and taking into account the
asymptotic oscillatory behavior of the functions $P$ and $Q$ at the
horizon ($x\to +\infty$) \be P_n^{i w/H}(\tanh{x H}) \sim
\frac{1}{\Gamma(1-i w/H)}e^{iwx} \ , \ee \be Q_n^{iw/H}(\tanh{x H})
\sim A(w) e^{iwx} + B(w)e^{-iwx} \ , \ee with \be A(w)=
\frac{-i\pi}{4\Gamma(1-i w/H)}(\coth\frac{\pi w}{2H} +
\tanh\frac{\pi w}{2H}) \ , \ee \be B(w)= \frac{-\pi}{4\Gamma(1-i
w/H)}\frac{\coth\frac{\pi w}{2H}}{ \sinh^2 \frac{\pi w}{2H}} \ , \ee
we find  that $|N_n(w)|^2 \equiv\frac{1}{w}|\tilde{N}_n(w/H)|^2$,
where $|\tilde{N}_n(w/H)|^2$ is the dimensionless function \be
\label{normalization} |\tilde{N}_n(w/H)|^2= \frac{1}{4 \pi  }
\frac{|\Gamma(1-iw/H)|^2}{|1+ \frac{i \pi}{4} \alpha_n(w)
(\coth{\frac{\pi w}{2 H}}+\tanh{\frac{\pi w}{2 H}})|^2}  . \ee
Therefore, the form of the modes at the physically relevant point
$\tilde{r}=0$, can be written as   \be u^{C}_{w}(\tilde{r}=0)=
e^{-iw\tilde{t}}\frac{\tilde{N}_n(w/H)}{\sqrt{4 \pi w}} H
\beta_n(w/H) \ , \ee where $\beta_n(w/H)$ is a dimensionless
function given in the appendix. With this we obtain \be \langle 0_C|
\phi(t, \vec{x})\phi(t, \vec{x})|0_C\rangle =  \frac{\hbar H^2
}{4\pi}\int_{\frac{m}{\hbar}}^{\infty}\frac{dw }{w} |\tilde{N}_n|^2
|\beta_n|^2 \ee

Taking into account the relation, $w^2= \bar{k}^2 +
m^2\hbar^{-2}$, one gets
the following  spectrum of fluctuations \be
\label{psC}\Delta^2_{\phi}(\bar{k}; C)=\hbar \frac{H^2}{4 \pi}
\frac{\bar{k}^2}{\bar{k}^2 + m^2\hbar^{-2}}
|\tilde{N_n}(\bar{k}H^{-1})|^2 |\beta_n (\bar{k}H^{-1})|^2] \ . \ee
The amplitude of these fluctuations depends only on the physical comoving scale $\bar{k}$. The
difference $\Delta^2_{\phi}(\bar{k}; dS)- \Delta^2_{\phi}(\bar{k};
C)$, which is the proposed spectrum of this paper, seems to be
driven, at first sight, by $H^2$.
However, explicit evaluation of the above formulas unravels a
miraculous simplification of the right-hand side of (\ref{psC}) when
the mass $m$ goes to zero. In this case, $n=3/2$, $\alpha_{3/2}=
\frac{2i}{\pi}\tanh \frac{\pi w}{2H}$, $\beta_{3/2}=
(1+iw/H)/[\Gamma(1-iw/H)\sinh^2\frac{\pi w}{2H}]$, and the
normalization factor is \be |\tilde{N}_{3/2}(w/H)|^2 =
\frac{|\Gamma(1-iw/H)|^2}{ \pi } \sinh^2{\frac{\pi w}{2 H}} \ . \ee
We find that, irrespective of the scale $\bar{k}$, the amplitude of
fluctuations is identical for both quantum states
\be \Delta^2_{\phi_{m=0}}(\bar{k};
dS)= \frac{\hbar H^2}{4\pi^2}(1 + \frac{\bar{k}^2}{H^2})=
\Delta^2_{\phi_{m=0}}(\bar{k}; C) \ . \ee

This result has a major consequence, since it implies that the
proposed power spectrum \be \Delta^2_{\phi}(\bar{k}) \equiv
\Delta^2_{\phi}(\bar{k}; dS)- \Delta^2_{\phi}(\bar{k}; C) \ , \ee {\it is
now driven by a different physical scale}, namely, the mass of the
scalar (inflaton) field, instead of the Hubble constant:
$\Delta^2_{\phi}(\bar{k})\propto m^2$ for small $m^2$.

Let us now
estimate the behavior of the proposed power spectrum for the
nonzero mass case. One immediately obtains that
\begin{eqnarray}
\Delta^2_{\phi}(\bar{k}) &=&\hbar H^2\left[\frac{1
}{8\pi}|H^{(1)}_{n}(\bar{k}H^{-1})|^2-\right.  \\ &&
\left.\frac{1}{4 \pi} \frac{\bar{k}^2}{\bar{k}^2 + m^2\hbar^{-2}}
|\tilde{N_n}(\bar{k}H^{-1})|^2 |\beta_n (\bar{k}H^{-1})|^2] \right]
\nonumber
\end{eqnarray}
This spectrum is still nearly scale-free  for $m^2/H^2\hbar^{2} \ll
1$ and $\bar{k} \approx H$.
\begin{table}
\begin{tabular}{| c| c | c | c | }
$\frac{m^2}{H^2\hbar^2}$ &  $10^{-1}$ &  $10^{-3}$ &  $10^{-5}$ \\  \hline
$\frac{\Delta_{\phi}^2(\bar{k})}{\Delta_{\phi}^2(\bar{k}, dS)}$ & $0.2212 \times10^{-1}$ & $0.2525 \times 10^{-3}$ & $0.2529 \times 10^{-5}$ \\ \hline
\end{tabular}
\caption{Ratio of the proposed power spectrum
$\Delta_{\phi}^2(\bar{k})$ by the standard value
$\Delta_{\phi}^2(\bar{k}, dS)$ at $\bar{k}=H$. }
\label{tab:spectra}
\end{table}
In Table \ref{tab:spectra} we compare the proposed power spectrum
$\Delta_{\phi}^2(\bar{k})$ with the standard spectrum
$\Delta_{\phi}^2(\bar{k}, dS)$ at $\bar{k}=H$ for different values
of the inflaton mass. We observe that the amplitude of the proposed
power spectrum scales with $m^2$ and the ratio with the
conventional spectrum for $m^2/(H\hbar)^2\leq 10^{-2}$ can be
approximated by
\begin{equation}\label{eq:ratio}
\left.\frac{\Delta_{\phi}^2(\bar{k})}{\Delta_{\phi}^2(\bar{k},dS)}\right|_{\bar{k}=H}\approx
0.25 \frac{m^2}{H^2\hbar^{2}} \ .
\end{equation}
This shows that our proposal for subtracting the amplitude of the
fluctuations of the comoving vacuum to define the power spectrum
produces a drastic reduction of its amplitude
provided that  $m^2$ is chosen sufficiently
small. It is worth noting that one of us \cite{parker07} already
found a similar behavior for the power spectrum on grounds of
adiabatic regularization of the two-point function. The fact of
getting similar numerical estimates from different approaches
supports the robustness of this result.

In this paper we have explored an alternative
definition for the power spectrum of quantum fluctuations in an
inflationary de Sitter universe.
An important result is that the amplitude of
quantum fluctuations for the de Sitter invariant and the comoving vacuum states in de Sitter space coincide
exactly in the massless case. This has major physical consequences.
The proposed spectrum is no longer driven  by the Hubble constant, but
instead by the effective mass of the scalar field. This provides a natural
way out of the problem of getting too large a magnitude for the amplitude
of inflaton fluctuations, since the magnitude can be automatically reduced
by several orders of magnitude, and it merits further exploration.

Furthermore, as pointed out by one of us in \cite{parker07},
the vanishing amplitude for the case of $m=0$ implies that
the tensor perturbations of the gravitational metric
during exponential inflation may be $0$. This is because
in the Lifshitz gauge, the two polarization components
of the gravitational tensor perturbations each satisfy the same
equation as a minimally-coupled scalar field with $m=0$.  It is
believed that the amplitude of the tensor perturbations is a gauge
invariant quantity. Then, our proposal, and the one in \cite{parker07},
would imply that the tensor to scalar ratio may be much smaller than
previously predicted. The standard predictions for this ratio
may soon come within the range of measurement.\\

This work has been partially supported by grants
FIS2005-05736-C03-03 and EU network MRTN-CT-2004-005104. L.Parker
has been partially supported by NSF grants PHY-0071044 and
PHY-0503366. I.A. and G.O. thank MICINN for a FPU grant and a JdC
contract, respectively. G.O. has also been supported by Perimeter
Institute for Theoretical Physics. Research at Perimeter Institute
is supported by the Government of Canada through Industry Canada and
by the Province of Ontario through the Ministry of Research \&
Innovation.\\

{\it Appendix.} Defining $\mu_{n}=1/2+n+iz$ we have
 \bea & &\beta_n(z)=
 \frac{i 2^{-n}}{\pi^2}
\frac{\cos{[\pi (n-i z)]}}{\cosh{\pi z}} \{i z
|\Gamma(\frac{\mu_{-n}}{2})|^2\Gamma(\mu_{n})+\nonumber
\\&+&\pi(2-\frac{m^2}{H^2\hbar^2}) [-\csc{[\frac{\pi}{2}\mu^*_{-n}]}
\frac{\Gamma(\mu_{-n})}{\Gamma(2+i z)} \Gamma(\mu_n) \times
\nonumber \\ &&  _2F_1(3/2-n,1/2-n+i z;2+i z;-1)+\nonumber\\&+&
\pi\frac{\csc{[\frac{\pi}{2}\mu_{-n}]}}{\Gamma(2-i z)}\sec{[\pi(n+i
z)}])]\times \nonumber\\& \times&
_2F_1(3/2-n,1/2-n-i z;2-i z;-1)]\}\times \nonumber\\
&\times& (\cos{[\pi\frac{n+i z}{2}]}+\sin{[\pi\frac{n+i z}{2}]}).
\nonumber \eea

\end{document}